\documentclass[aps,prl,reprint,superscriptaddress]{revtex4-1}
\usepackage{amsmath}
\usepackage{amsfonts}
\usepackage{graphicx}
\usepackage[usenames]{xcolor}
\usepackage{natbib}
\begin{document}

\title{Detection-Enhanced Steady State Entanglement with Ions}

\author{C D B Bentley}
\affiliation{Department of Quantum Science, Research School of Physics and Engineering, Australian National University, Canberra, Australia}

\author{A R R Carvalho}
\affiliation{Department of Quantum Science, Research School of Physics and Engineering, Australian National University, Canberra, Australia}
\affiliation{ARC Centre for Quantum Computation and Communication Technology, The Australian National University, ACT 0200, Australia}

\author{D Kielpinski}
\affiliation{Centre for Quantum Dynamics, Griffith University, Brisbane, Australia}

\author{J J Hope}
\affiliation{Department of Quantum Science, Research School of Physics and Engineering, Australian National University, Canberra, Australia}

\date{\today}

\begin{abstract}
Driven dissipative steady state entanglement schemes take advantage of coupling to the environment to robustly prepare highly entangled states.  We present a scheme for two trapped ions to generate a maximally entangled steady state with fidelity above 0.99, appropriate for use in quantum protocols.  Furthermore, we extend the scheme by introducing detection of our dissipation process, significantly enhancing the fidelity.  Our scheme is robust to anomalous heating and requires no sympathetic cooling.

\end{abstract}

\maketitle

Decoherence is notorious for being a major obstacle to the development of quantum technologies that require sustained quantum coherence. Typically, the interaction of a quantum system with its surrounding environment drives the system towards states with no traces of quantum behaviour. However, it has long been recognised that this does not have to be the case: by combining the irreversible decoherence dynamics with suitably chosen Hamiltonian evolution, one can design artificial reservoirs to steer the system to 
the desired stationary states. This quantum {\it reservoir engineering} method~\cite{Poya96PRL} has been used to investigate decoherence~\cite{Myat00Nat,Turch00PRA} and to design robust non-classical states~\cite{Carv01PRL} in the motional degrees of freedom of trapped ions. 

More recently, the interest in designing dissipative processes has been renewed with proposals to engineer many-body quantum states~\cite{Dieh08NatP,Vers09NP} and even perform quantum computation~\cite{Vers09NP}. Engineered reservoirs are robust to preexisting natural decoherence sources~\cite{Carv01PRL} and also to variations of parameters and initial conditions.  This has led to a number of proposals to prepare and stabilise quantum steady-states in a variety of systems \cite{Cho11PRL,Berm13PRL}, including cavity quantum electrodynamics~\cite{Piel07PRL,Sarl11PRL}, optomechanical systems~\cite{Toma12PRA,Wang13PRL}, and superconducting qubits~\cite{Reit13PRA}. Although experimentally challenging, engineered dissipation has been used to generate entanglement in atomic ensembles~\cite{Krau11PRL}, to implement quantum operations in ion traps~\cite{Barr11Nat}, and more recently to prepare Bell states in superconducting qubits~\cite{Shan13N} and in ion traps~\cite{Lin13N}.

Despite the intrinsic robustness of dissipative driven dynamics, the steady-state fidelities achieved in recent trapped ion experiments are far below the high fidelities achieved with more traditional time-dependent entangling gates~\cite{Benh08NP} and, therefore, still much lower than the fidelities required for a quantum information processing (QIP) system~\cite{Benh08NP,Knill05N}. In~\cite{Barr11Nat}, for example, the entangling mechanism relies on successive applications of quantum gates to generate a quantum operation. Even though a fidelity of $0.91$  has been achieved for a single cycle, fidelity decreases as the dynamics approach the continuous dissipative master equation limit. In contrast, Lin et al.~\cite{Lin13N} used continuous, time-independent fields and achieved fidelities of up to 0.75 that could be boosted to 0.89 using stepwise application of laser fields. In this case, the fidelity is limited by the intrinsic loss mechanisms present in the particular continuous driving scheme adopted.

In this paper we propose a steady-state scheme with fidelity above 0.99, which is further enhanced by photodetection.  Driven dissipation continuously pumps the system towards an asymmetric Bell steady-state, which is dark to the system dynamics.  The dominant loss mechanism is anomalous heating of the motional modes, reducing our fidelity by less than $0.01$ for current experimental rates.  Our scheme does not use sympathetic cooling, required in~\cite{Lin13N}, making it significantly simpler to implement.

We enhance our scheme by combining the dissipative state preparation with the detection of photons. Improvement in state generation through conditional dynamics has been explored in schemes relying on feedback~\cite{Carv08PRA, Stevenson2011,Stev11PRA} and also used recently in~\cite{Shan13N}. Here we show that by conditioning the system dynamics to the detection of photons spontaneously emitted into the environment, we obtain a significant fidelity enhancement with detection efficiencies of $10\%$.

\begin{figure}[tb]
     \centerline{\includegraphics[width=1\columnwidth]{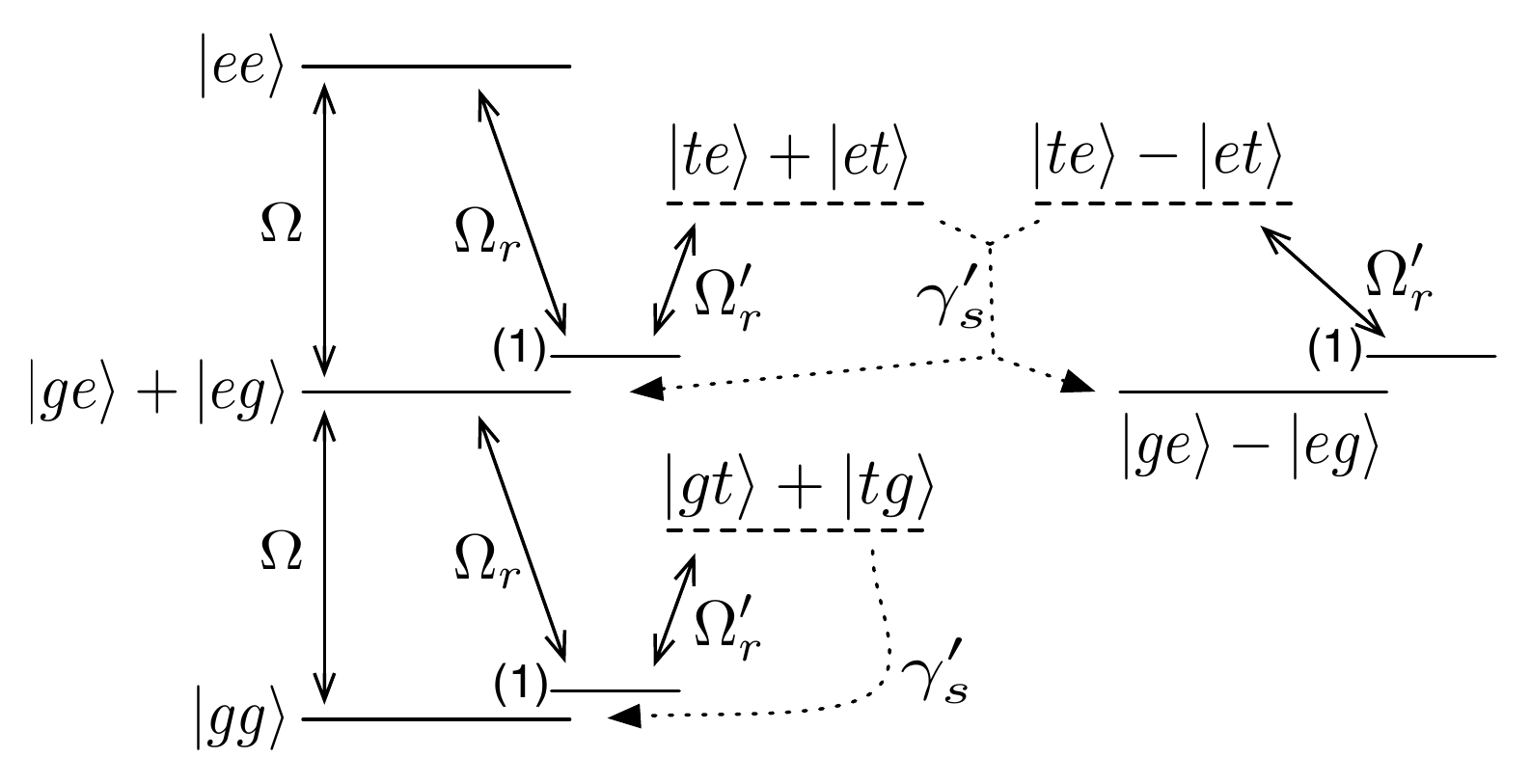}}
   \caption{Level scheme for the model with metastable levels $|g\rangle$ and $|e\rangle$, and temporary (short-lived) level $|t\rangle$.  The state $|ab\rangle$ represents the tensor product $|a\rangle \otimes |b\rangle$ of the individual ion states.  The first motional sideband is marked by (1).  Note the spontaneous emission from level $|t\rangle$ marked by dotted arrows.  The superpositions involving $|te\rangle$ both decay to an equal classical mixture of the symmetric and antisymmetric Bell states.  Motional heating and the much slower spontaneous emission from $|e\rangle$ are omitted for clarity.}
   \label{fig:full}
\end{figure}

Our level scheme, shown in Figure \ref{fig:full}, demonstrates the entanglement mechanism.  Lasers addressing both ions drive the carrier frequency at Rabi (angular) frequency $\Omega$ between ground ($|g\rangle$) and excited ($|e\rangle$) metastable states, couple the internal states to a motional mode by driving the red sideband at frequency $\Omega_r$, and excite the atoms from the ground to temporary level ($|t\rangle$) on the red sideband at frequency $\Omega_r'$.  The temporary level decays at rate $\gamma_s'$.  The result of a decay event is an equal mixture of the Bell symmetric and antisymmetric states, assuming that the ions remain indistinguishable upon any detection of the outgoing photon.  The antisymmetric (internal), ground (motional) state is a dark state of the system as there are no available transitions for the red sideband lasers, and the carrier transition laser jointly illuminates both ions, causing destructive interference to cancel any dynamics for this state.  In contrast, the symmetric component is driven back to the temporary level, providing increasing population in our target antisymmetric state.  We omit the motional heating channel in Figure \ref{fig:full} for clarity.  Heating populates the excited motional sidebands which in turn couple out of the dark internal state, the dominant loss mechanism.

We couple to red sidebands both for motional cooling and to keep the motional ground state of the antisymmetric state dark to the temporary level coupling.  The system dynamics are described by the master equation:
\begin{align}
\dot{\rho} &= \Lambda (\rho)  - i \Omega_r'[((b_1 + b_2) a^\dagger + (b_1^\dagger + b_2^\dagger) a),\rho] \nonumber \\
& \quad + \gamma_s' (\mathcal{D}[b_1]\rho + \mathcal{D}[b_2]\rho), \label{megrp} \\
\Lambda (\rho) &\equiv -i \Omega [(J_+ + J_-), \rho] - i \Omega_r[(J_- a^\dagger + J_+ a), \rho] \nonumber \\
& \quad +  \sum_i \gamma_s \mathcal{D}[\sigma_-^i]\rho + h_r (\mathcal{D}[a]\rho + \mathcal{D}[a^\dagger]\rho)
\end{align}
where $J_+ = \sigma_1^+ + \sigma_2^+$ and $J_- = \sigma_1^- + \sigma_2^-$, and where $\sigma_i^{+(-)}$ (de)excites atom $i$ between the ground and excited metastable levels.  The annihilation operator $a$ acts on the motional mode, and $b_i$ describes the $|t\rangle \rightarrow |g\rangle$ transition for atom $i$.  The metastable lifetime of the atoms is $\gamma_s$, the lifetime of the temporary level is $\gamma_s'$, and $h_r$ is the trap anomalous heating rate.  Decay and heating are represented by the decoherence superoperator $\mathcal{D}[A] \rho= A\rho A^\dagger - \frac{1}{2}(A^\dagger A \rho + \rho A^\dagger A)$.  The usual heating rate decoherence terms have prefactors based on bath temperature ($\bar{n}$ and $(\bar{n}+1)$), but in the limit of large $\bar{n}$, the prefactors both become $h_r$ (number of phonons gained per second).

Typical trapped ion species such as $^{40}$Ca$^+$ have suitable level structures for our scheme.  We take the $S_{1/2}$ level to be $|g\rangle$, and the $^{40}$Ca$^+$ metastable level $D_{3/2}$ can be used as an excited qubit state $|e\rangle$ with a lifetime of around 1~s~\cite{Gerr08EPJ}; we thus set our decay rate $\gamma_s = 1$~s$^{-1}$.
The $P_{3/2}$ level ($|t\rangle$) has decay rates on the order of $\gamma_s' = 10^8$~s$^{-1}$.  $\Omega_r'$ should be set such that $\gamma_s' \gg \Omega_r'/(2\pi)$ and the temporary level is negligibly populated, keeping the population in the qubit manifold.  We set $\Omega_r'=10^6$~rad/s in the following analysis.  For the specific $^{40}$Ca$^+$ case, no such single-photon transition exists, however a two-photon transition coupling to the $P_{3/2}$ ($|t\rangle$) level via the $D_{5/2}$ level would be effective, leaving the $D_{3/2}$ ($|e\rangle$) qubit state untouched.  Here off-resonant excitations marginally reduce the fidelity as quantified later in the Letter.

For $\gamma_s'$ much greater than the other evolution rates, we can adiabatically eliminate the temporary level to simplify calculation:
\begin{align}
\dot{\rho} &=  \Lambda (\rho) + \frac{4 \Omega_r'^2}{\gamma_s'} (\mathcal{D}[|g\rangle_1 \langle g| a]\rho + \mathcal{D}[|g\rangle_2 \langle g| a]\rho), \label{eq:adme}
\end{align}
where $|g\rangle_i \langle g|$ is the ground state projector for atom $i$.  This projector arises from the ground to temporary level coupling, leaving the excited state invariant.



\begin{figure}[tb]
     \centerline{\includegraphics[width=1\columnwidth]{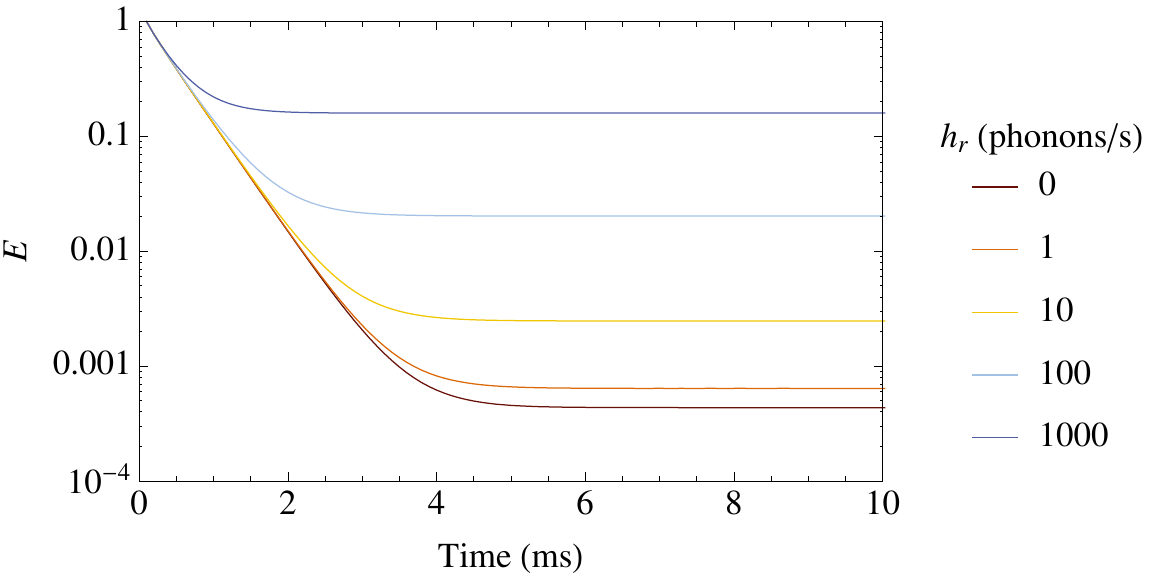}}
   \caption{Asymptotic error ($E = 1 - \text{Fidelity}$) for different anomalous heating rates $h_r$.  Error is plotted over time for $\Omega_r = 20$~krad/s, $\Omega=26$~krad/s, $\gamma_s=1$~s$^{-1}$ for different heating rates.}
   \label{fig:convt}
\end{figure}

The master equation gives the dynamics of the Bell antisymmetric state population shown in Figure \ref{fig:convt} for a $|gg\rangle$ initial state, using the coupling values $\Omega_r = 20$~krad/s and $\Omega=26$~krad/s.  Fidelity is defined as the population in the Bell antisymmetric state, and infidelity $E=1-\text{Fidelity}$.  After several milliseconds, the system converges to an asymptotic high-fidelity Bell state.  Anomalous heating varies depending on trap specifics, however heating rates below $h_r=10$~phonons/s have been achieved \cite{Roos99PRL}.  This provides a fidelity around 0.997, while 100~phonons/s gives a fidelity of 0.980.  Simulations were performed using the XMDS package \cite{Denn13CPC,xmdsweb}.

The master equation analysis follows the dynamics of a system without a measurement process.  A key component of driven dissipative schemes like ours is the spontaneous emission from a temporary level.  The system is driven towards a spontaneous emission event whenever it is not in the dark state: a detection event means that our fidelity is low.  Conversely, a lack of detection events indicates that our system is in the dark state.  Not detecting emissions thus heralds a high fidelity Bell antisymmetric state, an improvement over our unconditional steady state fidelity.  Detecting spontaneous emission events is challenging; for the following analysis we consider a photodetector collecting $\xi = 10\%$ of emissions.  

A stochastic master equation \cite{Carmichael1993,Molm93JOSAB} represents a particular trajectory of the system with a particular detection record.  The equation describes the evolution of the conditional density matrix $\rho_C$ for this detection record.  Poissonian noise $dN$ provides random spontaneous emission times interspersed with continuous evolution.  After adiabatic elimination for large $\gamma_s'$, the stochastic master equation is

\begin{align}
d\rho_C &= \sum_i dN_i \left( \frac{J_-^i \rho J_-^{i\dagger}}{\text{Tr} [J_-^{i\dagger} J_-^i \rho]} - \rho \right) + dt \{ \Lambda (\rho) \nonumber \\
&\quad + \frac{4 \Omega_r'^2}{\gamma_s'} [(1-\xi) \mathcal{D}[J_-^i]\rho + \xi ( -\frac{1}{2} (J_-^{i\dagger} J_-^i \rho \nonumber \\
&\quad + \rho J_-^{i\dagger} J_-^i) + \text{Tr} [ J_-^{i\dagger} J_-^i \rho] \rho  ) ] \}
, \label{eq:sadme}
\end{align}

where
\begin{align}
J_-^i &= |g\rangle_i \langle g| a, \\
\mathbb{E}[dN_i] &= dt \text{Tr} [|g\rangle_i \langle g| a^\dagger a \rho] \frac{4 \xi \Omega_r'^2}{\gamma_s'}.
\end{align}

\begin{figure}[tb]
     \centerline{\includegraphics[width=0.8 \columnwidth]{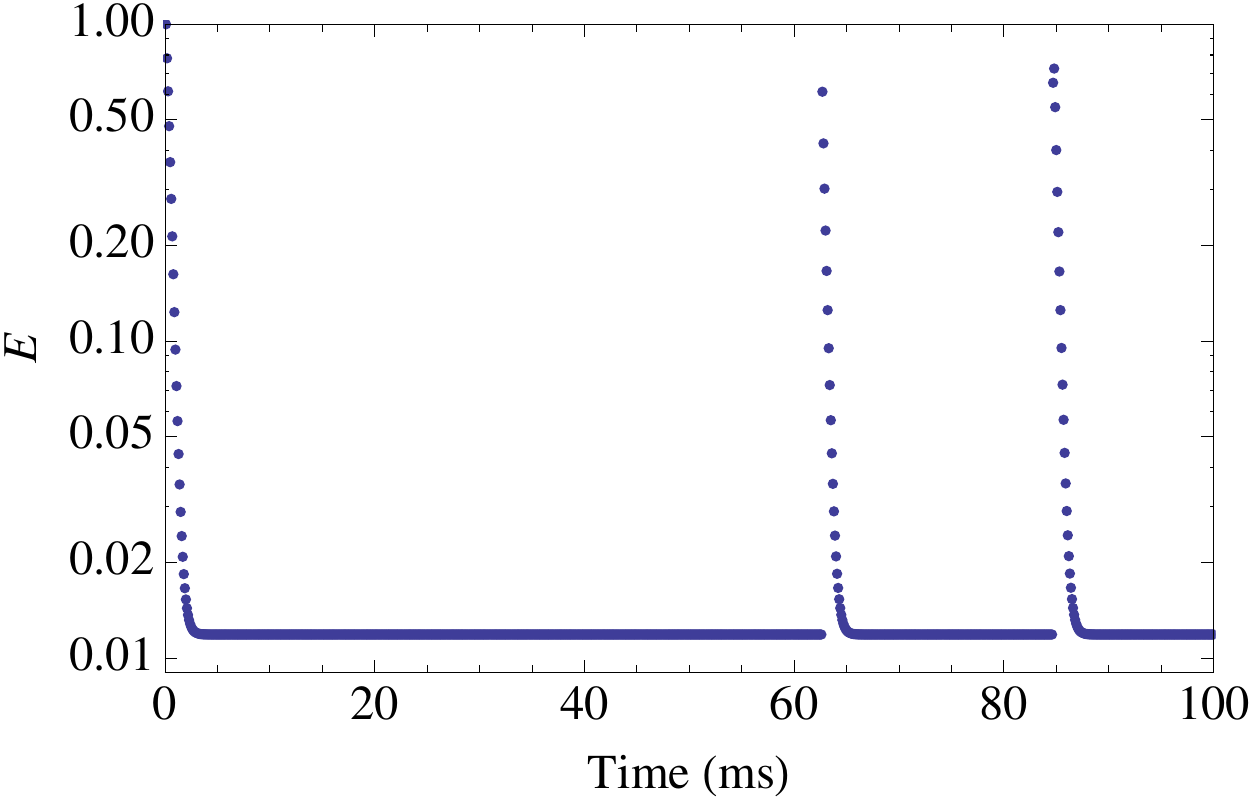}}
   \caption{Asymmetric Bell state error over time for a particular trajectory of equation (\ref{eq:sadme}) for $h_r = 100$~phonons/s, $\Omega_r = 20$~krad/s, $\Omega = 26$~krad/s.  The detection efficiency $\xi = 10\%$.}
   \label{fig:traj}
\end{figure}

Averaging this equation for different noise records returns the unconditional master equation, equation (\ref{eq:adme}).  Trajectories corresponding to particular measurement results were determined, as shown in Figure \ref{fig:traj}.  Spontaneous emission detection events cause the drops in fidelity, which are interspersed by convergence to a steady fidelity value.  The ion emission events are assumed to be indistinguishable.  Note that the choice of detected $J_-^i$ events could be replaced by (a)symmetric events $J_- = a(|g\rangle_1 \langle g| +(-) |g\rangle_2 \langle g|)$, which leads to the same asymptotes and negligible change in the convergence time.

\begin{figure}[tb]
     \centerline{\includegraphics[width=1\columnwidth]{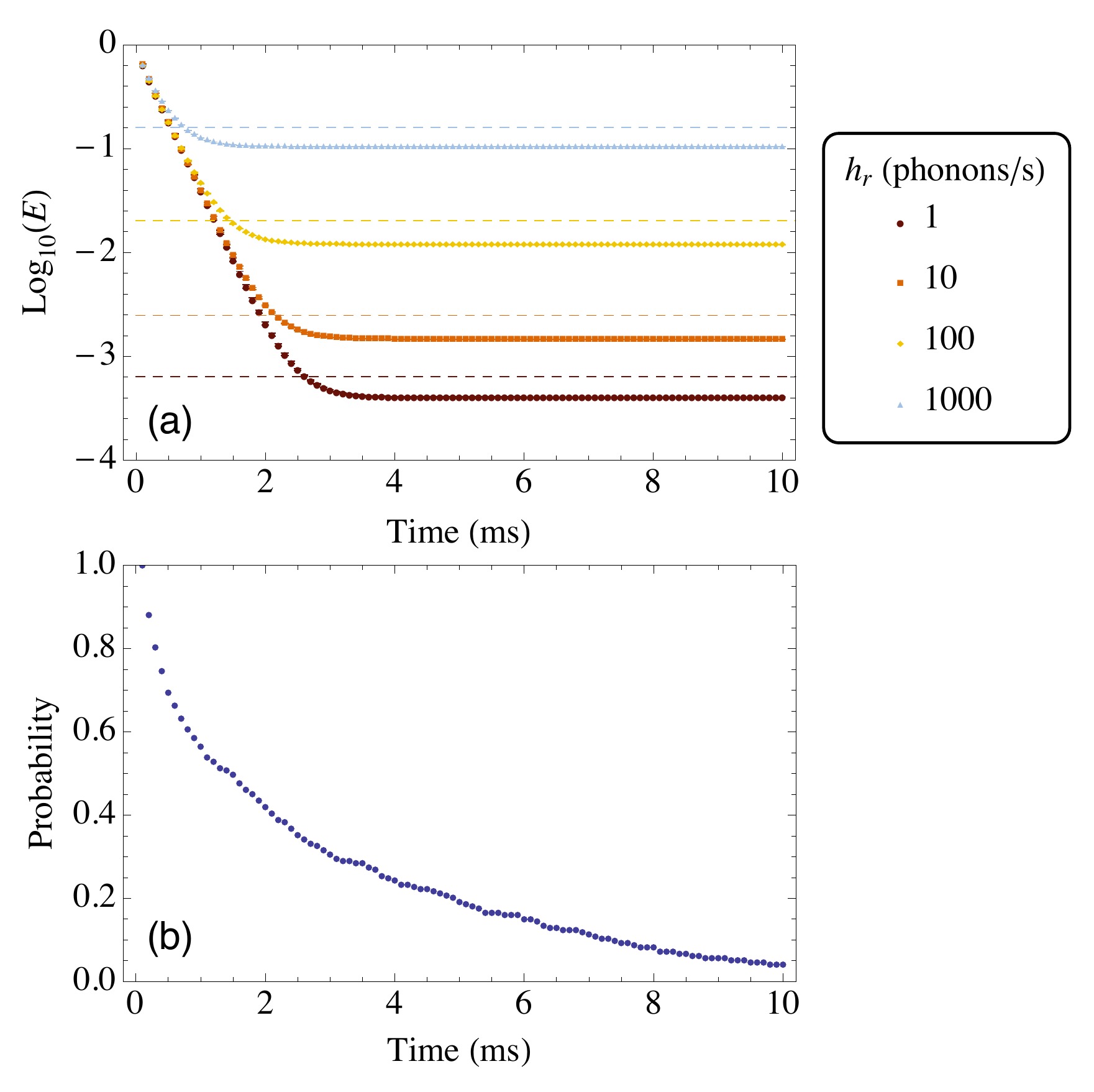}}
   \caption{For $\Omega_r = 20$~krad/s, $\Omega = 26$~krad/s, $\xi = 10\%$, (a): Log$_{10}$ of the asymmetric Bell state infidelity (error) after a detection event, conditioned on no subsequent detections for different heating rates.  Error bars mark the standard error, smaller than the data points.  The dashed lines represent the steady state infidelity with no photodetector.  (b): The probability of observing a single detection event and no further detections as a function of time for $h_r = 1000$~phonons/s.}
   \label{fig:condfid}
\end{figure}

Following a detection event, our fidelity is driven by the system to some high steady state value unless we observe another detection event.  Figure \ref{fig:condfid}(a) shows the infidelity after a detection event conditioned on no further detections.  Figure \ref{fig:condfid}(b) provides the probability that no subsequent detection will occur as a function of time, given a large heating rate $h_r = 1000$~phonons/s.  This probability is over 40\% at the fidelity asymptote of 0.895 for this heating rate.  For systems with less heating, reaching the asymptote becomes more probable.  Note also that if no detection event is observed since initialisation of the system, the same asymptotic convergence occurs.

The increase in fidelity given by introducing a measurement with $10\%$ detection is significant, while lower detection naturally gives less fidelity gain.  Table 1 explicitly shows the gain in fidelity for the conditional system with different detection efficiencies $F_C$($\xi$) over the unconditioned, measurement free system $F_U$.

\begin{table}[tb]
\begin{center}
  \begin{tabular}{ l c c c c }
    \hline
    $h_r$(phonons/s) & $F_U$ & $F_C$($1\%$) & $F_C$($3\%$) & $F_C$($10\%$) \\ \hline
    1 & 0.9994       & 0.9994 & 0.9995 & 0.9996         \\ 
    10 & 0.9975     & 0.9977 & 0.9979 & 0.9985       \\   
    100 & 0.9797   & 0.9810 & 0.9832 & 0.9881     \\
    1000 & 0.8404 & 0.8476 & 0.8607 & 0.8954   \\

  \end{tabular}
      \caption{Fidelity $F$ asymptotes for different heating rates, $h_r$.  $F_U$ is for the (unconditioned) system without measurement, which is contrasted with $F_C$, for the (conditional) system with $1\%$, $3\%$ and $10\%$ spontaneous emission detection.  These values are for $\Omega_r = 20$~krad/s, $\Omega=26$~krad/s, $\gamma_s = 1$~s$^{-1}$. }
\end{center} 
  \end{table}



\begin{figure}[tb]
     \centerline{\includegraphics[width=1\columnwidth]{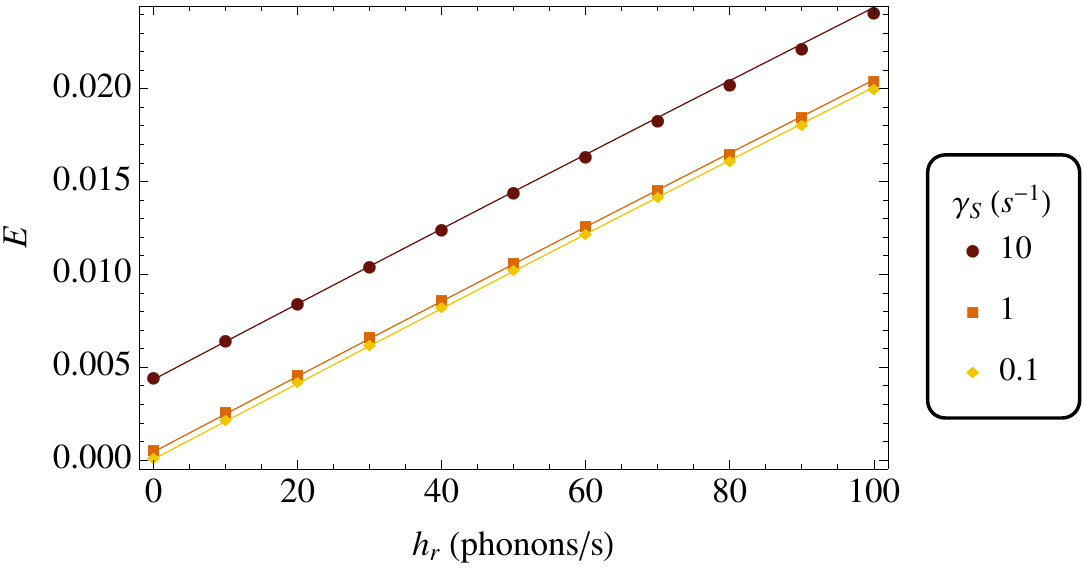}}
   \caption{The scheme is robust to noise: the steady state error as a function of anomalous heating rates $h_r$ and metastable excited state lifetime $\gamma_s$.  Here $\Omega_r = 20$~krad/s, $\Omega = 26$~krad/s.  The fit is a simple model based on steady state population transfer rates.}
   \label{fig:error}
\end{figure}

The model includes the effects of anomalous heating and spontaneous emission events, and we characterise the loss of fidelity that these processes introduce for the steady state scheme in Figure \ref{fig:error}.  The robust nature of the scheme is highlighted, as the infidelity $E$ roughly increases with the first order of the heating rate.  Metastable lifetimes above one second are readily achievable, thus the error from spontaneous emission is negligible.  

\begin{figure}[tb]
     \centerline{\includegraphics[width=0.9\columnwidth]{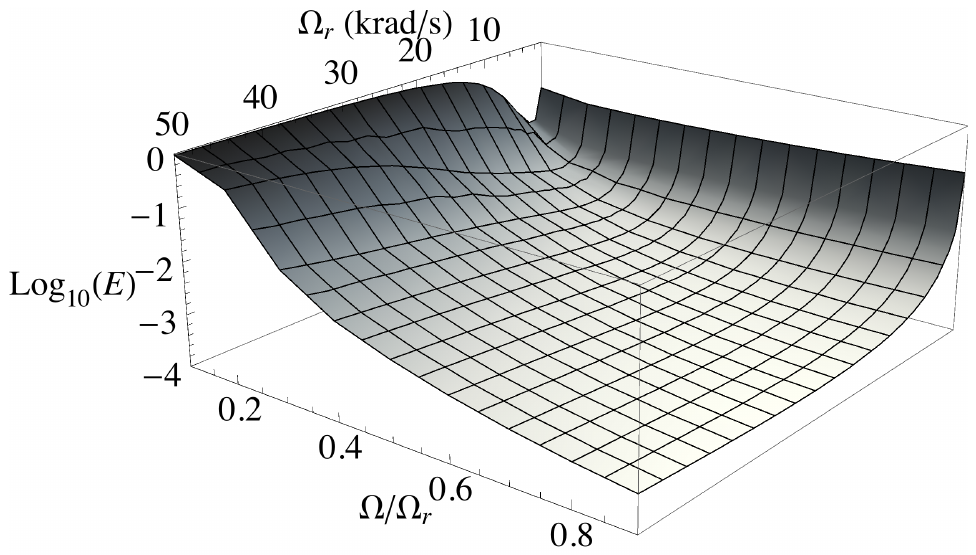}}
   \caption{Log$_{10}$ of the steady state error for different energy level coupling values, $\Omega$ and $\Omega_r$.  The scheme performs well, with $E<10^{-3}$ for a range of values.  Low values of $\Omega/\Omega_r$ (below around 0.2) do not reach the error asymptote before our imposed time cap of 0.1~s, giving the sharp error rise in this region.  High $\Omega/\Omega_r$ leads to occupation of high motional levels.  We truncate at 20 motional levels and consider only parameters with population below $10^{-6}$ of the maximal level.  Here $\gamma_s = 1$~s$^{-1}$, and $h_r = 0$~phonons/s.}
   \label{fig:fidom}
\end{figure}

The effect of laser instability or frequency errors in the coupling parameters is also important.  The steady-state behaviour is resistant to fluctuations in laser intensity.  Figure \ref{fig:fidom} demonstrates the robustness to variation in the coupling parameters $\Omega$ and $\Omega_r$, where fidelity $\simeq 1$ is reached for a range of the coupling rates.  Note that the anomalous heating is assumed to be negligible for this analysis, and the metastable spontaneous decay is set to $\gamma_s = 1$~s$^{-1}$.

We have assumed a uniform decay from the temporary level to our ground computational level.  For $^{40}$Ca$^+$, the branching ratios to $S_{1/2}$, $D_{5/2}$ and $D_{3/2}$ from the $P_{3/2}$ temporary level are 0.9347(3), 0.0587(2) and 0.00661(4) respectively \cite{Gerr08EPJ}.  Using these ratios and $^{40}$Ca$^+$ trap parameters from \cite{Roos99PRL}, we find an unconditional fidelity of 0.989 and a conditional fidelity of 0.995.  
The dominant error in this system is from off-resonant excitations on the cooling transitions, which leads to loss from our target state.  The fidelity given is optimised to balance the cooling rate, $10^4$~phonons/s here, with off-resonant loss.  The temporary level decay to the $D_{5/2}$ and $D_{3/2}$ levels has little effect as this population continues to be pumped towards the target state.


In summary, we make use of driven dissipation, with dynamics chosen such that the Bell antisymmetric state is a dark state of our system.  We then introduce measurement of the spontaneous decay, which indicates when the system is not in the dark state.  Conversely, when no decay is observed, we note a significant increase in the (conditional) steady state fidelity; for our $^{40}$Ca$^+$ model with anomalous heating of 5.3~phonons/s \cite{Roos99PRL}, the fidelity is 0.995.  The effects of spontaneous emission are also incorporated in the scheme, and its steady-state nature provides robustness to noise in the lasers.

This work was supported by the Australian Research Council Centre of Excellence for Quantum Computation and Communication Technology (Project number CE110001027) (ARRC), Australian Research Council Future Fellowships (FT110100513) (DK) and (FT120100291) (JJH) as well as DP130101613 (DK, JJH, ARRC).

\bibliographystyle{apsrev4-1.bst} 
\bibliography{librarylett2}

\end{document}